\documentstyle[preprint,floats,epsfig,aps]{revtex}
\begin{document}
\def\DESepsf(#1 width #2){\epsfxsize=#2 \epsfbox{#1}}
\input{epsf}

%1:topic==================================================================
\begin{center}
 \vskip 15mm
{\large $B \to \eta(\eta') K(\pi)$ in the Standard Model with Flavor Symmetry}
 \vskip 15mm
 % \Large
Han-Kuei Fu, Xiao-Gang He, Yu-Kuo Hsiao\\
Physics Department, National Taiwan University, Taipei, Taiwan, R.O.C\\
\vskip 15mm
\end{center}
%abstract=================================================================
\begin{abstract}
The observed branching ratios for $B\to K \eta'$ decays are much larger than
factorization predictions in the Standard Model (SM).
Many proposals have been made to reconcile the
data and theoretical predictions.
In this paper we study these decays within the SM
using flavor U(3) symmetry.
If small annihilation amplitudes are neglected, one needs 11 hadronic
parameters to describe
$B\to PP$ decays where $P$ can be one of the $\pi$, $K$, $\eta$
and $\eta'$ nonet mesons. We find that existing data are consistent with
SM with flavor U(3) symmetry.
We also predict several
measurable branching ratios and CP asymmetries for
$B \to K (\pi) \eta(\eta')$, $\eta(\eta')\eta(\eta')$ decays.
Near future experiments can provide important tests for
the Standard Model with flavor U(3) symmetry.
\end{abstract}
%=========================================================================
\newpage

%=========================================================================
Experimental data from CLEO, BaBar and Belle\cite{0,1,2,3} have
measured branching ratios of $B \to K \eta'$ around $6\times 10^{-5}$ which
are substantially larger than theoretical calculations based on
naive factorization approximation in the Standard Model (SM)\cite{4}.
Although there are some improvements in
calculating the branching ratios in the last few years by using
QCD improved factorization method\cite{5},
there are still large uncertainties in calculating
the branching ratios for $B\to K\eta'$
because of issues related to $\eta_1 - \eta_8$ mixing and
QCD anomaly associated with $\eta_1$. There are also many
speculations about possible new physics beyond the SM in these
decays\cite{6}. Before any claim can be made about new physics, one must
study the SM contributions in all possible ways
to see if it is really inconsistent with experimental data.

In this paper we carry out a systematic study of $B\to K \eta'$, and
more generally processes of $B\to PP$ decays
with at least one of the $P$
to be $\eta (\eta')$ in the final states by using flavor symmetry in the SM.
This way one can relate different decays to predict unmeasured
branching ratios and
CP asymmetries. Drastic deviations between the predicted
relations and experimental data can provide information about the SM and
models beyond. Similar considerations based on $SU(3)$
have been applied to $B \to PP$ decays\cite{7},
and shown to be consistent with data \cite{8}.
If one sticks to flavor SU(3)
symmetry, one needs to introduce the singlet $\eta_1$ in the theory and
to add additional amplitudes to describe new decay modes\cite{quinn}.
One may also consider to promote the flavor SU(3) symmetry to flavor U(3)
symmetry such that $\eta_1$ is automatically included in the theory.

Flavor $U(3)$ symmetry has been studied in Kaon decays.
There there are non-negligible symmetry breaking effects. For
$B$ decays one may also expect symmetry breaking effects to exist.
There are also some studies of U(3) symmetry for $B$ decays\cite{9}.
Present data, however, are not able to make clear statement about
whether this symmetry is badly broken. In this paper we will
take the flavor U(3) symmetry as working hypothesis and to study whether
experimental data can be explained by carrying out a systematic analysis.
We find that the SM with flavor U(3) symmetry can explain all existing data,
in particular can obtain large branching ratios for $B\to K\eta'$ decays.
We also predict some unmeasured branching ratios and CP asymmetries which
can be used to further test the theory.

The quark level effective Hamiltonian can be written as\cite{10}
\begin{eqnarray}
 H_{eff}^q = {G_{F} \over \sqrt{2}} [V_{ub}V^{*}_{uq} (c_1 O_1 +c_2 O_2)
   - \sum_{i=3}^{11}(V_{ub}V^{*}_{uq}c_{i}^{uc} +V_{tb}V_{tq}^*
   c_i^{tc})O_{i}].
\end{eqnarray}
Here $V_{ij}$ are KM matrix elements.
The coefficients $c_{1,2}$ and $c_i^{jk}$ are the Wilson Coefficients
which  have been evaluated by several groups\cite{10} with
$|c_{1,2}|>> |c_i^{jk}|$.
$O_i$ are operators consist of quarks and gluons.

The $B\to PP$ decay amplitudes can be parameterized as
\begin{eqnarray}
A(B\to PP) = <PP|H_{eff}^q|B> = {G_F\over \sqrt{2}}[V_{ub}V^*_{uq} T
+ V_{tb}V^*_{tq}P],
\end{eqnarray}
where $B = (B_u,  B_d,  B_s) = (B^-, \bar B^0, \bar B^0_s)$ which form a
fundamental representation of $SU(3)$ (or ($U(3)$).
The amplitudes $T$ and $P$ are related to the hadronic matrix elements
$<PP|O_i|B>$ which are very difficult to calculate. For our purpose, however,
we only need to know the fact that under $SU(3)$ (or $U(3)$)
$O_{1,2}$, $O_{3-6,11}$, and $O_{7-10}$
transform as $\bar 3 + \bar 3' +6 + \overline {15}$,
$\bar 3$, and $\bar 3 + \bar 3'
+6 + \overline {15}$, respectively\cite{7}, and to parameterize the amplitudes
according to SU(3) (or U(3)) invariant amplitudes to be discussed below.

As mentioned earlier that there are two approaches to the problem from
flavor symmetry point of view.
We first work with the U(3) symmetry approach.
In this case, the $\pi$, $K$, $\eta_8$ and $\eta_1$ form a nonet representation
of $U(3)$, $M^i_j$, with

\begin{eqnarray}
(M_i^j)
&=&
\left(\begin{array}{ccc}
\frac{\pi^0}{\sqrt{2}}+ \frac{\eta_8}{\sqrt{6}} &\pi^+& K^+  \\
\pi^-&-\frac{\pi^0}{\sqrt{2}}+ \frac{\eta_8}{\sqrt{6}}  &K^0  \\
K^- & \bar K^0 &    -2 \frac{\eta_8}{\sqrt{6}}
\end{array} \right) +
\left(\begin{array}{ccc}
\frac{1}{\sqrt{3}}\eta_1&0&0\\
0&\frac{1}{\sqrt{3}}\eta_1&0\\
0&0&\frac{1}{\sqrt{3}}\eta_1
\end{array} \right)
\nonumber
\end{eqnarray}

One can write the
$T$ amplitude for $B \to PP$ in terms of the flavor U(3) invariant amplitudes
as
\begin{eqnarray}
T&=& A_{\bar 3}^TB_i H(\bar 3)^i (M_l^k M_k^l)
+ C^T_{\bar 3}B_i M^i_kM^k_jH(\bar 3)^j \nonumber\\
&+&\tilde A^T_{6}B_i H(6)^{ij}_k M^l_jM^k_l
+ \tilde C^T_{6}B_iM^i_jH(6)^{jk}_lM^l_k\nonumber\\
&+&A^T_{\overline{15}}B_i H(\overline{15})^{ij}_k M^l_jM^k_l
+C^T_{\overline{15}}B_iM^i_jH(\overline{15} )^{jk}_lM^l_k\nonumber\\
&+&B^T_{\bar 3}B_i H(\bar 3)^i M^j_j M^k_k
+\tilde B^T_{6}B_i H(6)^{ij}_k M^k_j M^l_l\nonumber\\
&+&B^T_{\overline{15}}B_i H(\overline{15})^{ij}_k M^k_j M^l_l
+D^T_{\bar 3}B_i M^i_jH(\bar 3)^j  M^l_l\;.
\label{amp}
\end{eqnarray}

In Table \ref{su3tb} we list all decay amplitudes involving $\eta_{1,8}$.
The amplitudes containing only $K$ and $\pi$ in the final states can be found in
Ref. \cite{8}.
There are a few new features for the
U(3) amplitudes in eq. (\ref{amp})
compared with the SU(3) amplitudes for $B\to PP$.
The last four terms are new. In SU(3) case because the traceless
condition of $M_{i}^j$, these terms are automatically zero.
With SU(3) symmetry, the amplitudes $\tilde A_6$ and $\tilde C_6$
always appear in the
combination of $\tilde C_6 - \tilde A_6$\cite{7}. This degeneracy is, naively
lifted in processes with $\eta_1$ in the final states. It seems that
there is the need of having both $\tilde C_6$ and $\tilde A_6$ to
describe the decays increasing the total number of hadronic parameters by one.
However, this is not true since that the $\tilde
A^T_6$ and $\tilde C^T_6$ terms
in decay modes with $\eta_1$ in the final state can be written
as $C^T_6 =
\tilde C^T_6 - \tilde A^T_6$, and the additional $\tilde A^T_6$ be absorbed by
redefining the amplitude $B^T_6 = \tilde B_6^T + \tilde A^T_6$.
In Table I we therefore have listed the decay amplitudes in terms of
the independent U(3) invariant amplitudes, $C^T_{\bar 3, 6, \overline{15}}$,
$A^T_{\bar 3, \overline{15}}$, $B^T_{\bar 3, 6, \overline{15}}$ and
$D^T_{\bar 3}$.

We now describe the other approach to include $\eta_1$ in $B \to PP$ decays.
Here one treats $\eta_1$ as a singlet of $SU(3)$ and parameterizes
the decay amplitudes according to SU(3) symmetry.
In this case there are also additional four new terms,

\begin{eqnarray}
T_{new} &=& a^T B_i H(\bar 3)^i \eta_1\eta_1 +
b^T B_i M^i_j(8) H(\bar 3)^j \eta_1\nonumber\\
&+&  c^TB_i H(6)^{ik}_lM^l_k(8) \eta_1
+ d^T B_i H(\overline {15})^{ik}_l M^l_k(8)\eta_1.
\end{eqnarray}
In the U(3) symmetry limit, we have

\begin{eqnarray}
&&a^T = A^T_{\bar 3} + 3 B^T_{\bar 3} + {1\over 3} C_{\bar 3}^T + D_{\bar 3}^T,
\;\;\;\;\;\;b^T = {2\over \sqrt{3}} C^T_{3} + \sqrt{3} D^T_{\bar 3},\nonumber\\
&&c^T= {2\over \sqrt{3}} \tilde A_{\bar6}^T+\sqrt{3} \tilde
B^T_{\bar6}+{1\over\sqrt{3}} \tilde C^T_{\bar 6},
\;\;\;\;d^T =
{2\over \sqrt{3}} A^T_{\overline{15}}+\sqrt{3} B^T_{\overline{15}} +{1\over\sqrt{3}} C^T_{\overline{15}}.
\end{eqnarray}
We also see from the above that one can use $C_6^T$ and $B^T_6$
to absorb
$\tilde A^T_6$ by writing,
$c^T = \sqrt{3} B^T_6 + C^T_6$.
It is interesting to note that both approaches discussed above
introduce the same number of new parameters, four of them, into the theory.
In our analysis we will work with the flavor U(3)
symmetry.

To obtain the amplitudes for $B$ decays with at least one $\eta (\eta')$
in the final states, one also needs to
consider $\eta - \eta'$ mixing,
\begin{eqnarray}
\left(\begin{array}{c}
\eta\\
\eta'
\end{array}\right)
=
\left(\begin{array}{cc}
cos\theta&-sin\theta\\
sin\theta&cos\theta
\end{array}\right)
\left(\begin{array}{c}
\eta_8\\
\eta_1
\end{array}\right)\;.
\end{eqnarray}
The average value of the mixing angle $\theta$ is $-15.5^\circ \pm
1.3^{\;\circ}$\cite{11}. We will use $\theta = -15.5^\circ$ in our fit.

%fitting
\normalsize

There are similar $U(3)$ invariant amplitudes for the
penguin contributions. We indicate them as $C^{P}_{\bar 3,6,\overline{15}}$,
$A^{P}_{\bar 3,\overline{15}}$, $B^{P}_{\bar 3,6,\overline{15}}$ and
$D^{P}_{\bar 3, 6,\overline{15}}$. The amplitudes
$A_i$ and $B_i$ are referred as annihilation amplitudes because the $B$ mesons
are first annihilated by the interaction Hamiltonian and two light
mesons are then created. In total there are
22 complex hadronic parameters (44 real parameters with one of them to
be an overall unphysical phase).
However simplification can be made because the following relations in the SM,
\begin{eqnarray}
C^P_6(B^P_6) &=& - {3\over 2}{c_9^{tc} - c_{10}^{tc}\over c_1-c_2-3(c_9^{uc}-c_{10}^{uc})/2}
C^T_6(B^T_6) \approx - 0.013 C^T_6(B^T_6) \;,\nonumber\\
C^P_{\overline {15}}(A^P_{\overline {15}},B^P_{\overline {15}}) &=& -{3\over 2}
{c_9^{tc}+c_{10}^{tc}\over c_1+c_2-3(c^{uc}_9
+c_{10}^{uc})/2} C^T_{\overline {15}} (A^T_{\overline {15}},B^T_{\overline {15}})
\approx +0.015 C^T_{\overline {15}} (A^T_{\overline {15}},B^T_{\overline {15}}).
\label{P2T}
\end{eqnarray}
Here we have used the Wilson Coefficients obtained in Ref.\cite{10}.

We comment that in finite order perturbative calculations
the above relations are
renormalization scheme and scale dependent.
One should use a renormalization scheme consistently. We
have checked with different renormalization schemes and find that numerically
the changes are less than 15\% for different schemes.
In obtaining the above relations, we have also neglected small contributions
from $c_{7,8}$ which cause less than 1\% deviations.

Using the above relations the number of independent hadronic
parameters are reduced
which we choose to be, $C_{\bar 3}^{T,P}(A^{T,P}_{\bar 3})$, $C_6^T$,
$C^T_{\overline {15}}(A^T_{\overline {15}})$, $B^{T,P}_{\bar3}$,
$B^{T}_{\bar6}$, $B^{T}_{\overline{15}}$, $D^{T,P}_{\bar3}$.
An overall phase can be removed without loss of generality,
we will set $C^P_{\bar 3}$ to be real. There are in fact only 25
real independent
parameters for $B\to PP$ in the SM with flavor U(3) symmetry,\\[2 mm]
\hspace*{2 cm}
$C_{\bar 3}^P,\;\;C_{\bar 3}^T e^{i\delta_{\bar 3}},\;\;
C^{T}_6e^{i\delta_6},\;\;
C^{T}_{\overline{15}}e^{i\delta_{\overline{15}}},\;\;
A^T_{\bar 3}e^{i\delta_{A^T_{\bar 3}}},\;\;
A^P_{\bar 3} e^{i\delta_{A^P_{\bar 3}}},\;\;
A^T_{\overline{15}} e^{i\delta_{A^T_{\overline{15}}}},\;\;\\
\hspace*{2 cm}
B^{T}_{\bar3}e^{i\delta_{B^T_{\bar 3}}},\;\;
B^{P}_{\bar3}e^{i\delta_{B^P_{\bar 3}}},\;\;
B^{T}_{\bar6}e^{i\delta_{B^T_{\bar 6}}},\;\;
B^{T}_{\overline{15}}e^{i\delta_{B^T_{\overline{15}}}},\;\;
D^{T}_{\bar3}e^{i\delta_{D^T_{\bar 3}}},\;\;
D^{P}_{\bar3}e^{i\delta_{D^P_{\bar 3}}}.$
\\[2 mm]
Further the amplitudes $A_i$ and $B_i$ correspond to annihilation contributions and
are expected to be small which is also supported by data\cite{8}.
If the annihilation amplitudes are neglected,
there are only 11 independent hadronic parameters
\begin{eqnarray}
C_{\bar 3}^P,\;\;C_{\bar 3}^T e^{i\delta_{\bar 3}},\;\;
C^{T}_6e^{i\delta_6},\;\;
C^{T}_{\overline{15}}e^{i\delta_{\overline{15}}},\;\;
D^{T}_{\bar3}e^{i\delta_{D^T_{\bar 3}}},\;\;
D^{P}_{\bar3}e^{i\delta_{D^P_{\bar 3}}}.
\label{ci}
\end{eqnarray}
The phases in the above can be defined in such a way that all $C_i^{T,P}$
and $D^{T,P}_i$
are real positive numbers.

\begin{table}%[htb]
\caption{U(3) decay amplitudes for $B\to PP$ with at least one of the $P$
being a $\eta_8$ or $\eta_1$.\label{su3tb} }
\footnotesize
\begin{eqnarray}
\begin{array}{l}
\hspace{-3mm}
\left.
\begin{array}{l}
\Delta S = 0\\
T^{B_u}_{\pi^- \eta_8}(d)={2\over \sqrt{6}}
(C^T_{\bar 3}  - C^T_6 + 3 A^T_{\overline {15}} + 3C^T_{\overline {15}}),\\
T^{B_d}_{\pi^0 \eta_8}(d)= {1\over \sqrt{3}}
(-C^T_{\bar 3}  + C^T_6 + 5 A^T_{\overline {15}} + C^T_{\overline {15}}),\\
T^{B_d}_{\eta_8 \eta_8}(d)={1\over \sqrt{2}}
(2A^T_{\bar 3} + {1\over 3} C^T_{\bar 3} - C^T_6
-A^T_{\overline {15}} + C^T_{\overline {15}}),\\
T^{B_s}_{K^0 \eta_8}(d)=
-{1\over \sqrt{6}}(C^T_{\bar 3}  + C^T_6 -  A^T_{\overline {15}}
-5 C^T_{\overline {15}}),\\
T^{B_u}_{\pi^- \eta_1}(d)={2\over \sqrt{3}}
(2 C^T_{\bar 3}  + C^T_6 + 6 A^T_{\overline {15}} + 3C^T_{\overline
{15}}\\\;\;\;\;\;\;\;\;\;\;\;\;\;\;\;\;\;\;\;\;\;\;
+ 3 B^T_6+ 9 B^T_{\overline {15}}+3 D^T_{\bar 3}),\\
T^{B_d}_{\pi^0 \eta_1}(d)={1\over \sqrt{6}}
(2 C^T_{\bar 3}  + C^T_6 -10 A^T_{\overline {15}} -5 C^T_{\overline {15}}\\
\;\;\;\;\;\;\;\;\;\;\;\;\;\;\;\;\;\;\;\;\;\;
+ 3 B^T_6- 15 B^T_{\overline {15}}+3 D^T_{\bar 3}),\\
T^{B_d}_{\eta_1 \eta_8}(d)={1\over 3\sqrt{2}}
(2 C^T_{\bar 3}  -3 C^T_6 + 6 A^T_{\overline {15}} + 3C^T_{\overline {15}}\\
\;\;\;\;\;\;\;\;\;\;\;\;\;\;\;\;\;\;\;\;\;\;
- 9 B^T_6 + 9 B^T_{\overline {15}}+3 D^T_{\bar 3}),\\
T^{B_d}_{\eta_1 \eta_1}(d)={\sqrt{2}\over 3}
(3 A^T_{\bar 3} + C^T_{\bar 3} + 9 B^T_3 +3 D^T_{\bar 3}),\\
T^{B_s}_{K^0 \eta_1}(d)={1\over \sqrt{3}}
(2 C^T_{\bar 3}  - C^T_6 - 2 A^T_{\overline {15}} -  C^T_{\overline {15}}\\
\;\;\;\;\;\;\;\;\;\;\;\;\;\;\;\;\;\;\;\;\;\;
- 3 B^T_6 - 3 B^T_{\overline {15}}+3 D^T_{\bar 3}),\\
\end{array}
\right.
\left.
\begin{array}{l}
\Delta S = -1\\
T^{B_u}_{\eta_8K^-}(s)= {1\over\sqrt{6}}(-C^T_{\bar 3}
   + C^T_{6} - 3A^T_{\overline {15}} +9 C^T_{\overline {15}}),\\
T^{B_d}_{\eta_8 \bar K^0}(s)=  -{1\over \sqrt{6}} (C^T_{\bar 3}
+ C^T_{6} - A^T_{\overline {15} } -5 C^T_{\overline {15} }),\\
T^{B_s}_{\pi^0\eta_8}(s)= {2\over \sqrt{3}}( C^T_{6}
+2 A^T_{\overline {15}} - 2C^T_{\overline {15}}),\\
T^{B_s}_{\eta_8\eta_8}(s)= \sqrt{2}(A^T_{\bar 3} +{2\over 3} C^T_{\bar 3}
- A^T_{\overline {15} } - 2C^T_{\overline {15}})\;,\\
T^{B_u}_{K^- \eta_1}(s)={1\over \sqrt{3}}
(2 C^T_{\bar 3}  + C^T_6 + 6 A^T_{\overline {15}} + 3  C^T_{\overline {15}}\\
\;\;\;\;\;\;\;\;\;\;\;\;\;\;\;\;\;\;\;\;\;\;
+ 3 B^T_6+ B^T_{\overline {15}}+3 D^T_{\bar 3}),\\
T^{B_d}_{\bar K^0 \eta_1}(s)={1\over \sqrt{3}}
(2 C^T_{\bar 3}  - C^T_6 -2 A^T_{\overline {15}} -  C^T_{\overline {15}}\\
\;\;\;\;\;\;\;\;\;\;\;\;\;\;\;\;\;\;\;\;\;\;
- 3 B^T_6 -3 B^T_{\overline {15}}+3 D^T_{\bar 3}),\\
T^{B_s}_{\pi^0 \eta_1}(s)={2\over \sqrt{6}}
( C^T_6 -4 A^T_{\overline {15}} - 2 C^T_{\overline {15}}\\
\;\;\;\;\;\;\;\;\;\;\;\;\;\;\;\;\;\;\;\;\;\;
- 3 B^T_6 +6 B^T_{\overline {15}}),\\
T^{B_s}_{\eta_1 \eta_8}(s)={-\sqrt{2}\over 3}
(2 C^T_{\bar 3} -6 A^T_{\overline {15}} - 3  C^T_{\overline {15}}\\
\;\;\;\;\;\;\;\;\;\;\;\;\;\;\;\;\;\;\;\;\;\;
- 9 B^T_{\overline {15}}+3 D^T_{\bar 3}),\\
T^{B_s}_{\eta_1 \eta_1}(s)={\sqrt{2}\over 3}
(3 A^T_{\bar 3}+C^T_{\bar 3} + 9 B^T_3+ 3 D^T_{\bar 3}),\\
\end{array}
\right.
\end{array}
\nonumber
\end{eqnarray}
%\normalsize
\end{table}
At present many $B \to PP$ decay modes have been measured
at B-factories \cite{1,2,3}.
It is tempting to use experimental data to fix all the hadronic parameters
described earlier. It has been shown that if processes involving
$\eta (\eta')$
are not included, it is indeed possible to determine all the
SU(3) invariant amplitudes, $A_i$ and $C_i$\cite{7}.
When processes involving $\eta (\eta')$ are also included,
a meaningful determination of all hadronic parameters
(25 of them) is, however, not possible at present because of too many
parameters.
We therefore in the following
neglect the annihilation amplitudes, which are anticipated to be small,
to see if all data can be
reasonably explained, in particular to see if large $B \to K \eta'$
branching ratios can be obtained,
with only 11 parameters given in eq. (\ref{ci}).
This is a nontrivial task.
Remarkably we find that all data can, indeed, be
well explained.

We use the averaged CLEO, BaBar and Belle data\cite{1,2,3}
shown in Table \ref{asy1} and
\ref{asy2}
to fix the unknown 11 hadronic parameters by carrying out a global $\chi^2$
analysis. The results are shown in Table \ref{DD}.
In our analysis due to the lack of knowledge of
the error correlations from experiments, in obtaining the averaged
error bars, we have, for simplicity, taken them to be uncorrelated and assumed
to obey Gaussian distribution taking the larger one between
$\sigma_+$ and $\sigma_-$ to be on the conservative side.
Experimental data on
$\epsilon_K$, $B-\bar B$ mixing, $|V_{cb}|$, $|V_{ub}/V_{cb}|$, and
$\sin2 \beta$ provide very stringent constraints on the KM matrix elements
involved in our analysis\cite{8,12,13}.
We have treated them in our analysis as known parameters
with the values $\lambda = 0.2196$, $A = 0.854$, $\rho = 0.25$ and
$\eta = 0.34$ ($\gamma = 53.4^\circ$) determined from the most recent data.

\begin{table}[htb]
\caption{The best fit values and their 68\% C.L. ranges for the
hadronic parameters.}\label{DD}
\begin{tabular}{|c|c|c|}
            &   central value   &   error range         \\  \hline

$ C^P_{\bar 3} $ & 0.136 & 0.003 \\ \hline
$ C^T_{\bar 3} $ & 0.174 & 0.090 \\ \hline
$ C^T_6 $ & 0.244 & 0.077 \\ \hline
$ C^T_{\overline{15}} $ & 0.147 & 0.011 \\ \hline
$ \delta_{  \bar 3} $ & $ 85.6 ^\circ $ & $ 29.8 ^\circ $ \\ \hline
$ \delta_ 6 $ & $ 79.0 ^\circ $ & $ 17.4 ^\circ $ \\ \hline
$ \delta_{\overline{15}} $ & $ 8.9 ^\circ $ & $ 15.3 ^\circ $ \\ \hline \hline
$ D^P_{\bar 3} $ & 0.122 & 0.011 \\ \hline
$ D^T_{\bar 3} $ & 0.940 & 0.340 \\ \hline
$ \delta_{D^P_{\bar 3}} $ & $ -85.0 ^\circ $ & $ 6.0 ^\circ $ \\ \hline
$ \delta_{D^T_{\bar 3}} $ & $ -83.7 ^\circ $ & $ 16.5 ^\circ $ \\ \hline
\end{tabular}
\end{table}

Using the above determined hadronic parameters, we study several
other unmeasured branching ratios and
CP violating rate asymmetries $A_{CP}$ for $B \to PP$ defined by,
\begin{eqnarray}
A_{CP} = {\Gamma( B_i \to PP) - \Gamma(\bar B_i \to \bar P \bar P)\over
\Gamma(B_i\to PP) + \Gamma(\bar B_i\to \bar P\bar P)}.
\end{eqnarray}
The results are shown
in Tables \ref{asy1} and \ref{asy2}.

\footnotesize
\begin{table}[htb]
\caption{The central values and 68\% C.L. allowed ranges for
branching ratios (in units of
$10^{-6}$) and CP asymmetries for processes with no $\eta$
or $\eta'$ in the final states.} \label{asy1}
\begin{tabular}{|c|c|c|c|c|}
    &   \multicolumn{2}{c}{Branching Ratios}\vline                          &   \multicolumn{2}{c}{CP Asymmetries}\vline                         \\  \hline
    &   Experiment   &   Fit   &  Experiment   &  Fit   \\  \hline
$B_u \to \pi^- \pi^0$  &       $5.6 \pm 0.9$   & $ 5.5 ^{+ 0.9 }_{ -0.9 } $ & $0.06\pm 0.16$ & $ 0.00 $ \\ \hline
$B_u \to K^- K^0$   &       $-0.6\pm 0.8$ & $ 0.8 ^{+ 0.4 }_{ -0.2 } $ & & $ -0.49 ^{+ 0.84 }_{ -0.46 } $ \\ \hline
$B_d \to \pi^+ \pi^-$  &          $4.8\pm 0.5$ & $ 4.7 ^{+ 0.5 }_{ -0.5 } $ &      $0.51\pm 0.19$                    & $ 0.45 ^{+ 0.11 }_{ -0.12 } $ \\ \hline
$B_d \to \pi^0 \pi^0$  &       $2.0\pm 0.8$        & $ 1.9 ^{+ 0.8 }_{ -0.7 } $ & & $ 0.27 ^{+ 0.18 }_{ -0.33 } $ \\ \hline
$B_d \to \bar K^0 K^0$  & & $ 0.7 ^{+ 0.4 }_{ -0.2 } $ & & $ -0.49 ^{+ 0.84 }_{ -0.46 } $ \\ \hline
$B_u \to \pi^- \bar K^0$    &       $18.2\pm 1.7$ & $ 20.1 ^{+ 1.1 }_{ -1.1 } $ &    $0.04\pm 0.08$ & $ 0.02 ^{+ 0.03 }_{ -0.04 } $ \\ \hline
$B_u \to \pi^0 K^-$   &     $12.9\pm 1.2$ & $ 10.8 ^{+ 0.6 }_{ -1.2 } $ &     $-0.10\pm 0.08$ & $ -0.01 ^{+ 0.06 }_{ -0.07 } $ \\ \hline
$B_d \to \pi^+ K^-$   &      $18.5 \pm 1.0$ & $ 18.9 ^{+ 0.9 }_{ -0.4 } $ &     $-0.09\pm 0.04$ & $ -0.11 ^{+ 0.03 }_{ -0.03 } $ \\ \hline
$B_d \to \pi^0 \bar K^0$ &      $10.3\pm1.5$ & $ 8.9 ^{+ 0.4 }_{ -0.5 } $ & $0.03\pm 0.37$ & $ -0.06 ^{+ 0.06 }_{ -0.06 } $ \\ \hline
$B_s \to K^+ \pi^-$      & & $ 4.4 ^{+ 0.5 }_{ -0.5 } $ & & $ 0.45 ^{+ 0.11 }_{ -0.12 } $ \\ \hline
$B_s \to K^0 \pi^0$   & & $ 1.8 ^{+ 0.7 }_{ -0.7 } $ & & $ 0.27 ^{+ 0.18 }_{ -0.33 } $ \\ \hline
$B_s \to K^+K^- $   & & $ 17.8 ^{+ 0.8 }_{ -0.4 } $ & & $ -0.11 ^{+ 0.03 }_{ -0.03 } $ \\ \hline
$B_s \to K^0 \bar K^0$        & & $ 17.7 ^{+ 0.5 }_{ -1.0 } $ & & $ 0.02 ^{+ 0.03 }_{ -0.04 } $ \\
\end{tabular}
\end{table}

\footnotesize
\begin{table}%[htb]
\caption{The central values and their 68\% C.L. allowed ranges for
branching ratios (in units of
$10^{-6}$) and CP asymmetries with at least one of the
final mesons to be a $\eta$ or $\eta'$.}\label{br2} \label{asy2}
\begin{tabular}{|c|c|c|c|c|}
    &   \multicolumn{2}{c}{Branching Ratios}\vline                          &   \multicolumn{2}{c}{CP Asymmetries}\vline                         \\  \hline
    &  Experiment    &   Fit   &  Experiment   &   Fit   \\  \hline

$B_u \to \pi^- \eta$     & $4.1\pm 0.9$ & $ 4.1 ^{+ 0.9 }_{ -0.9 } $ & $-0.51\pm 0.20$ & $ -0.23 ^{+ 0.14 }_{ -0.14 } $ \\ \hline
$B_u \to K^- \eta$     & $3.1\pm 0.7$ & $ 3.3 ^{+ 0.6 }_{ -0.4 } $ & $-0.32\pm 0.22$ & $ -0.37 ^{+ 0.08 }_{ -0.09 } $ \\ \hline
$B_u \to K^- \eta'$     & $77.6\pm 4.8$ & $ 72.8 ^{+ 3.9 }_{ -3.8 } $ & $0.04\pm 0.04$ & $ 0.07 ^{+ 0.04 }_{ -0.04 } $ \\ \hline
$B_d \to \bar K^0 \eta$     & $2.6\pm 0.9$ & $ 2.4 ^{+ 0.5 }_{ -0.6 } $ & & $ -0.21 ^{+ 0.07 }_{ -0.09 } $ \\ \hline
$B_d \to \bar K^0 \eta'$     & $58.3\pm 6.0$ & $ 66.5 ^{+ 3.7 }_{ -3.6 } $ & $0.08\pm 0.16$ & $ 0.12 ^{+ 0.04 }_{ -0.04 } $ \\ \hline
$B_u \to \pi^- \eta'$     & & $ 16.8 ^{+ 16.0 }_{ -\;\;9.7 } $ & & $ -0.18 ^{+ 0.15 }_{ -0.09 } $ \\ \hline
$B_d \to \pi^0 \eta$     & & $ 1.2 ^{+ 0.6 }_{ -0.4 } $ & & $ -0.94 ^{+ 0.15 }_{ -0.03 } $ \\ \hline
$B_d \to \pi^0 \eta'$     & & $ 7.8 ^{+ 3.8 }_{ -4.3 } $ & & $ -0.38 ^{+ 0.19 }_{ -0.35 } $ \\ \hline
$B_d \to \eta \eta$     & & $ 3.1 ^{+ 1.3 }_{ -1.1 } $ & & $ -0.33 ^{+ 0.10 }_{ -0.13 } $ \\ \hline
$B_d \to \eta \eta'$     & & $ 7.6 ^{+ 5.3 }_{ -3.4 } $ & & $ -0.20 ^{+ 0.12 }_{ -0.20 } $ \\ \hline
$B_d \to \eta' \eta'$     & & $ 5.4 ^{+ 4.5 }_{ -3.1 } $ & & $ -0.11 ^{+ 0.14 }_{ -0.28 } $ \\ \hline
$B_s\to K \eta$     & & $ 2.8 ^{+ 1.5 }_{ -1.2 } $ & & $ 0.17 ^{+ 0.07 }_{ -0.07 } $ \\ \hline
$B_s \to K\eta'$     & & $ 19.1 ^{+ 7.0 }_{ -9.0 } $ & & $ -0.39 ^{+ 0.17 }_{ -0.26 } $ \\ \hline
$B_s \to \pi^0 \eta$     & & $ 0.05 ^{+ 0.10 }_{ -0.10 } $ & & $ 0.98 ^{+ 0.03 }_{ -0.14 } $ \\ \hline
$B_s \to \pi^0 \eta'$     & & $ 0.07 ^{+ 0.10 }_{ -0.10 } $ & & $ 0.91 ^{+ 0.09 }_{ -0.10 } $ \\ \hline
$B_s \to \eta \eta$     & & $ 7.0 ^{+ 1.5 }_{ -1.5 } $ & & $ -0.23 ^{+ 0.09 }_{ -0.09 } $ \\ \hline
$B_s \to \eta \eta'$     & & $ 24.1 ^{+ 1.4 }_{ -1.5 } $ & & $ 0.08 ^{+ 0.04 }_{ -0.04 } $ \\ \hline
$B_s \to \eta' \eta'$     & & $ 68.3 ^{+ 4.4 }_{ -4.5 } $ & & $ 0.09 ^{+ 0.05 }_{ -0.05 } $ \\
\end{tabular}
\end{table}

\normalsize

We now discuss some implications of the results obtained and draw conclusions.
We see from Tables \ref{asy1} and \ref{asy2}
that the best fit values for the known
branching ratios are in good agreements with data,
in particular large $B\to K \eta'$ can be obtained.
The minimal $\chi^2$ in our fit is
16 with total of 23 data points from $B\to PP$ decays
and the
11 hadronic parameters in eq.(\ref{ci}) as fitting parameters.
The value of 1.33 for the $\chi^2$ per degree of freedom
represents a reasonable fit.
In our fit the $\eta-\eta'$ mixing parameter $\theta$ is fixed at
the average value determined from other data\cite{11}.
We checked the sensitivity
of the final results on $\theta$ within the allowed region and find the changes
are small.
We also find that if one reduces the U(3) symmetry
to the SU(3) case, just fitting data on $B\to \pi\pi, \pi K, KK$,
the values obtained
for $C_i$ are not very much different than what obtained here. This indicates
that the parameters $C_i$ are stable when promoting $SU(3)$ to $U(3)$.
The large branching ratios for $B\to K \eta'$ are due to
the new parameters $D_{\bar 3}$.
The $U(3)$ assumption
can provide a good approximation for $B\to PP$ decays.
The Standard Model with flavor U(3) symmetry
is not in conflict with existing data.

In our fit, we did not include the branching ratios which only have
information on their upper bounds, such as $Br(B_d \to K^- K^+)$,
$Br(B_d \to \bar K^0 K^0)$ and
$Br(B_u \to \pi^- \eta')$. Since we neglected annihilation contributions,
the mode $B_d \to K^- K^+$ has vanishing branching ratio which is
consistent with data. The stringent upper limit\cite{1,2,3}
of $10^{-6}$ on $B_d\to K^-K^+$ branching ratio
supports the expectation that annihilation contributions are small.
The predicted branching ratio of
$0.7\times 10^{-6}$ for $B_d \to \bar K^0 K^0$
is safely below the experimental bound.
The predicted branching ratio for $Br(B_u \to \pi^-\eta')$ is
$16.8^{+16.0}_{-\;\;9.7} \times 10^{-6}$. The central value is slightly larger
than the 90\% C.L. allowed upper bound $12\times 10^{-6}$.
But the 68\% C.L. range is consistent
with data. At present it is too early to claim conflict of theory with
data. But this mode can be used to test the theory. Should
a branching ratio much smaller than the central value predicted
here be measured in the future, it is an indication
that the assumptions made need to be modified.

Using existing experimental data, we have determined 11 hadronic parameters
needed to describe $B\to PP$ decays with flavor U(3) symmetry.
We have compared with QCD improved
factorization calculations and found that the magnitudes of the parameters
are of the same order of magnitude. In our fit, we determined two
U(3) invariant amplitudes, $D^{T,P}_{\bar 3}$. These are particularly
difficult to estimate from theoretical calculations because these amplitudes
may be related to QCD anomalies. Also in factorization approach,
it is not possible to reliably
calculate the phases in the hadronic parameters. In our fit, we find that
these phases can be sizeable. Further improved theoretical method is needed
to have a better understanding of $B\to PP$ decays.

Using the hadronic parameters determined from existing data,
we have predicted several unmeasured branching ratios.
These predictions can be used to test the theory.
There are six modes involving at least one $\eta$ (or a $\eta'$) in the final
states for $B_d$ decays. Among them $B_d \to \pi^- \eta'$ has the largest
branching ratio,
it is a clear test for the theory. The other modes are also in the reach of
near future data from B-factories. There are also seven $B_s$ decay modes
with at least one $\eta$ or $\eta'$ in the final states.
Several of the branching ratios are predicted to be large, in particular
the predicted branching ratio for $B_s \to \eta' \eta'$ is about
$7\times 10^{-5}$ which can be measured at future hadron colliders and can
provide another crucial test for theory.

We have also obtained
interesting predictions for CP asymmetries in $B\to PP$ decay modes.
Many of the
predicted central values for the CP asymmetries are larger than $10\%$ which
can be measured in the near future. These modes can provide important
information about CP violation in the Standard Model.

In conclusion, we have carried out an systematic analysis for $B\to PP$ decays
in the SM with flavor U(3) symmetry. This approach allows one to study
$B$ decays involving
at least one $\eta$ or $\eta'$ in the final states. We find that
all existing data can be explained, in particular large branching ratios
for $B\to K \eta'$ are possible. There is no conflict between the Standard Model
and present experimental data.
We have also predicted several unmeasured
branching ratios and CP asymmetries within the reach of near
future B-factories. Future experimental data
will provide crucial information on flavor symmetries and also the
Standard Model.

\end{document}